IoT Enabled Insurance Ecosystem - Possibilities, Challenges and Risks

Jai Manral

NTT DATA FA Insurance Systems (NDFS)

Contact: jaimanral@mail.com

Author Note

Jai Manral, Consultant - Insurance Practice, Asia Business Unit, NTT DATA FA Insurance Systems (NDFS).

This research acknowledges the support of NDFS, a leading specialist IT solutions and services provider to the Insurance Industry.

Correspondence concerning this article should be addressed to Jai Manral, NTT DATA FA Insurance Systems (NDFS), Bangalore, 560060. Email: jaimanral@mail.com

Abstract

Internet of Thing (IoT) is looking over to overhaul the business processes of many industries including insurance domain. The current line of business such as Property and Casualty, Health, and Life Insurance can avail tremendous benefits from the contextual and relevant data being generated from billions of connected devices; Smartphone's, wearable gadget and other electronic smart sensors. For P&C insurer's the biggest challenges is not the rapidly changing environment but tackling these challenges with strategies linked to the past.

This paper presents the key opportunities, challenges, potential applications and risks of IoT in Insurance domain. It crystallomancy on the efficient use of data analytics from data generated in IoT ecosystem and highlights business model changes that Insurance industries may face in near future.

IoT Enabled Insurance Ecosystem - Possibilities, Challenges and Risks

## 1. The Connected Ecosystem

It is estimated that by 2025, IoT will be pervasive with data explosion from connected devices to an extent where a family of four can have over 100 connected devices [1]. This will change the perception and interaction of consumers between themselves and insurers. Not only this change will impact the core of P&C business model but will bring along new opportunities, a shift from restitution to prevention. In a connected environment where multiple interconnected devices gather real time information and provide meaningful insight to data taking automatic, tangible and collaborative preventive actions [2].

Smart devices capable of communicating certain events are advancements in home automation and enhancing security, which has already brought about a change in property insurance offerings. Telematics and Smart cars are enhancing the safety and security of both man and machine. Usage based insurance like Pay-As-You-Drive Auto (PAYD) and Pay-How-You-Drive (PHYD) plans are new customized products leveraging on these technologies. Health and Life insurance can avail benefits by offering customized products and pricing for their customers by utilizing volume of data generated from smart devices like wearable gadgets.

## 2. Internet of Things-An Overview

The Internet of Things refers to a network of identifiable physical objects (things) interconnected through wireless and wired connections and are able to interact and share information with each other. Integration of these smart objects with current infrastructure of internet is shaping the concept of IoT [3]. In recent years, cost reduction in hardware and network infrastructure with rapid development of wireless communication has surge the development of merging scenarios.

The idea of IoT has more than a decade of existence and was developed by the MIT Auto-ID Labs [3]. The use of this technology paired with RFID was mostly used in product tracking application for supply chain [4]. Over the years, IoT has evolved from the RFID centric approach to a vision of IoT society, where different objects are connected ubiquitously [3]. Internet Business Solutions Group (IBSG) at Cisco has supported this idea by terming IoT as a state or point in time when internet accessed by things will surpass their human counterparts [3]. This connection can be established by embedding transmitters and sensors into wide range of products or objects enabling them with unique identification and processing power allowing them to connect to the internet [4]. IoT as in such will not only allow things to communicate between them but will create new ways for humans to communicate with objects.

In a study by [5], IoT is believed to lead a third wave of Information Technology industry's revolution. This is very well pointed out by developed nations as the most important strategic pillar for driving economic and technology developments and innovations respectively. Cisco's Internet Business Solutions Group in its survey said 12.5 billion things are already connected to the internet in 2010 and is expected to rise by 50 billion by 2020 [5]. In an IoT network where the communication takes place between objects themselves and with humans, generate immense information which is readily accessible, this can be utilized in any value chain and resulting IoT applications will have the potential to change the business models in any industry [3] [32].

**2.1 Key Technologies of IoT**

We present here three IoT components, taxonomies of each component can be found in the study of [6] [7] [8]. The first component is the Hardware - made up of actuators, sensors and embedded communication hardware. Second component is Middleware - computing tools and on demand storage for processing data using data analytic techniques, and the third component is Presentation - the visualization and interpretation tools which can be designed and used on various applications [9]. Below we discuss few technologies which makes up three components as discussed.

### 2.1.1 Radio Frequency Identification (RFID)

IoT is based on the network technology where RFID (Radio Frequency Identification) is one of the key technologies. RFID system consists of two prominent components: electronic tags and reader. Where electronic tags are attached to objects which are to be identified and reader reads /write the information depending on the technology and configuration deployed [10]. These RFID tags are compact devices which can store limited data and can transmit the same when queried over RFID reader. This system works as an advanced non-contact automatic identification technology [11]. The passive class of RFID tags is widely used in retail and supply chain management. These tags are not battery powered but uses the power of RFID reader's interrogation signal for communication [11] [9].

### 2.1.2 Wireless Sensor Networks (WSN)

Over the recent years, innovation in wireless communications and development of low power integrated circuits with high efficiency and low cost has boomed its usage in remote sensing applications [6]. These factors have enabled to create sensor networks with large number of intelligent sensors enabling the collection, analysis, processing and dissemination of valuable information gathered across variety of environments [6]. The components that make for WSN monitoring networks includes:

a) WSN hardware - called as nodes consist of sensors, processing units, transceiver units and power supply

b) WNS communication stack - Nodes are mostly employed in an ad-hoc manner for most applications, where designing topology, routing and MAC layer is critical for longevity and scalability of WSN networks [6].

c) Middleware - Like Open Sensor Web Architecture (OSWA), the mechanism is provided to combine cyber infrastructure with a Service Oriented Architecture (SOA) and sensor networks for providing access to heterogeneous sensor resources [6].

d) Secure Data aggregation - It is required for extending the operation lifetime of the network and ensuring reliable data collected from sensors [27].

### 2.1.3 Addressing schemes

The foremost and important thing for IoT is to identify billion of objects uniquely and to control remote devices through internet. With the launch of IPv6 (Internet Protocol version 6) in the year 2012, assigning IP address to every object without any constrains had enabled the connectivity between millions of devices [14] alleviating the device identification problems. However, heterogeneous nature of wireless nodes, concurrent operations and the convergence of data from different devices aggravate the problems further [14] [30].

### 2.1.4 Data Storage and Mining

As we have discussed earlier, the communication between billion of devices in IoT architecture will tend to generate enough data which needs to be collected and stored for further analysis and mining. The issues with storage, ownership and expiry of data will become critical issues [13]. New artificial intelligence algorithms which can be developed which are intelligent enough to route the data in storage and be self-sufficient to automated decision making like handling ownership of data.

### 2.1.5 Visualization

For any IoT application, visualization is critical as this provides the interface for users to interact with the environment of connected objects. This will enable us to convert large chuck of data into meaningful information which can be used for analytics or decision making process. This constitutes both event detection and visualization of raw and modeled data which can be presented to end users as per their needs.

## 2.2 IoT Applications

Though there are hundreds of IoT applications identified and considered by different industries, we have logically categories them in two parts:

Applications Category One

This comprises the idea of connecting heterogeneous and interconnected devices tagged with unique IDs which can interact with other devices/machines, infrastructure and physical environment. Applications under this category are largely for Machine to Machine (M2M), Machine to Infrastructure (M2I) and Machine to Nature (M2N) communications. The IoT here plays a remote track, command, control and route (TCC&R) role [5] [6].

Applications Category Two

The second category is about identifying and working on the data which gets collected by the end nodes (from smart devices with collecting and transmitting capabilities) and are used for data mining across various businesses.

## 3. Delivering Insurance using IoT

Underwriting is the core of Insurance business. It deals with insurance pricing and premiums which are based on the rating parameters derived from historical loss data trends. In insurance business, law of large numbers and spread of associated risk is used to arrive at the pure premium for a pool of homogeneous risk. The main challenge faced by Underwriters is to calculate loss propensity and exposure at each risk level [21]. The process of Underwriting is initiated when a policy is sold, on the other hand risk changes over time. This is factored in UW on regular basis. Though insurance industries are data rich but their ability to monitor claims propensity at each exposure level and at each risk over the lifetime of a policy is limited. This approach results in premium being subsidized by good risks and taking the hit by bad risks [22].

IoT ecosystem makes it possible to generate real time data (captured from devices and sensors of insured objects and people) which than can be utilized for rating parameters. It will enable insurers to make personalized customer specific products based on their risk profiles. Underwriting will move from conventional risk assessment driven by historical data for a pool of risk to a personalized risk based pricing model. This will enable insurers to do premium revision on base of loss/risk conditions of insured during the policy period. This will transform how the products and services are currently offered by insurers into more customized customer centric and tailored products.

In an estimate, by 2025 IoT will have a significant impact on the business models of insurers be it P&C, Life or Health. With the new development in technologies and rolling out intelligent products like smart homes, cars and devices, which can be connected and able to exchange relevant information will create new opportunities for insurers. Google's pod shaped smart car prototypes provides a glimpse in the future [29]. These smart cars or smart homes can provide real time risk information which then can be used by insured and insurers to adopt necessary risk management techniques and thus mitigate losses [31]. Some of the biggest claims in homeowners' policy are due to perils such as water damage, theft, explosions and fire [26].

These can be recorded, transmitted and analyzed by using smart connected sensors capable of measuring temperature, smoke, humidity and other factors. Insurers can tie up with machine to machine (M2M) vendors with capabilities to monitor and alert policy holders on imminent risks.

State Farm, US biggest underwriter for homeowner insurance offers discounts to policy holders using Iris home automation system [24].

This will help insures to broaden their horizon from conventional attributes like previous claims histories and can leverage on connected technologies to provide data from within the prospect/insured property and thus accurately predict the loss making characteristics. This will also enable the insurers to assess the behavioral and moral hazard associated with insured [2].

Internet of Things opens the way to real-time and constant monitoring which can help in a claim process. Insurers will have better chance to identify the loss cause and even settle claims before being notified by insured [2] [25]. It will help in identifying and proactively response to fortuitous claims. The data recorded during the claim event can be useful for claim adjusters in accessing the degree and impact of losses. Auto data transmission from IoT devices will reduce the physical investigation, reducing loss adjustment expenses and speed up the claim settlement process, and hence improving insured overall claim settling experience.

<div align="center">Data Enrichment – Redefining Insurance Data Mining</div>

Businesses look for accurate and details picture of their clients or customers to mitigate risk, reduce fraud and increase profitability while enhancing customer services. It is thus vital for them to have reliable data to provide insights on customer habits. Insurers are not short of customer data, traditional or modern way of doing business has resulted in generating large chunks of data. However, this massive data with little intelligence have failed to extract measurable business insight for Insurance business to its full extent, making them "data rich and information poor".

Data Science provides great impact in business process from manufacturing to customer services. In a study by [12], the methodical approach of utilizing this data is discussed; collection, analysis and availability. First is to collect more external data, by third party / governmental agencies. Second is to use this data for predictive analysis. Finally, make relevant data accessible across company, for facilitating decision making.

The advent of Big Data with sophisticated analytics has relieved new prospects for Insurers to do business seminal and faster. Principal Financial Group, by leveraging on internal data was able to successfully automate Universal life and Retail variable UW business. Business rules coupled with predictive modeling have reduced their efforts and has made UW process faster. Their projection is to underwrite 40% of business without human intervention [12].

Usage based Insurance (UBI) motor insurers rely heavily on data aggregated by use of Telematics. This black box device records data, like driving speed, miles covered, frequent lane changes, hard breaking, cornering and over speeding. Data generated from these parameters provide insight on driving habits of individuals. Insurers use this data for risk assessments and hence calculating underwriting premiums. The biggest advantage of IoT devices in collecting data is its availability and remote access. Data when accessed in real time can shrink the gap between event and action time.

## 4. Challenges, Security and Risks

IoT device's possibilities are endless but the implementation is tricky. In order to transform their businesses leveraging on this connectedness insures must focus on security aspects of IoT. Businesses before investing on this technology should access the risks associated with this system connected to internet. As we are aware, any reliance on technology can cause disruption big or small, insurers must acknowledge these risks.

## 4.1 Data Privacy and Security

With boom in IoT enabled devices, insurers seeking customer data for calibrating there pricing for risk covers will have to make customers agreeable to share their personal data by assuring data privacy and security for shared information. The internet based connectivity of IoT architecture is prone to cyber-attacks. With rise in cyber breaches, Cyber-attacks stand as one of the biggest threats to business and governments across the globe. According to one estimate, cyber related crimes are costing 400 USD billions to businesses every year [19].

In 2014, a unit of baby monitor system developed by Focsam, Chinese based company, was hacked [20]. Later revealed, 40,000 out of 46,000 units were missing important security update. These security mishaps of hacking and accessing data can be adverted using various security techniques, like data hiding techniques [23]. In April 2015, the U.S. Government Accountability Office issued a report highlighting threats caused by increased interconnectedness between ground systems and airplanes [15].

Cyber-attacks can be carried out in IoT enabled devices compromising its security and then it may facilitate the attack on the network it is connect to. For example, a compromised IoT device can be used to launch denial of service attack [16]. An effective DOS attack can assemble a large number of connected devices. Compromised devices can also be used as to send malicious emails [28].

The vulnerability of life saving devices, such as insulin pumps or Telematics devices is a big concern [18]. If compromised, these devices can have fatal consequences. Though various researches [17] on securing the communication between IoT enabled devices has made significant contribution on IoT security aspects, insurance companies and IT service providers will have to be certain before rolling out the products and services.

## 4.2 Malfunctioning – The false alarm

One of the challenges for any IT service provider is to adhere the scenarios for device malfunctioning. The interconnection network of IoT devices where they exchange data has the risk that a malfunction could lead to system failure. Malfunction devices could feed in wrong information to other devices and this information up in the system can have adverse effects [18]. Crises like these can create havoc in connected networks.

In a sensor driven connected ecosystem loss mitigation depends on sensors and connected devices. The process of Subrogation (loss recovery), will have a significant change in how business is conducted presently, with the increase in number of connected cars and homes in future. However, in case a device malfunctions, it will be difficult to identify negligent party. For example, if a sensor fails to respond, consequently smart device fails to take corrective measure and there could be a fault from the policy holder or the supplier which resulted in a loss. In this scenario, it will be difficult for insurer to find the degree of fault from parties involved and establish recovery [2].

## 5. Capitalizing on Opportunities

The success of insurer depends on taking advantages of emerging future technologies and revamps their market strategies and business process. Insures can use IoT to differentiate them in this increasingly commoditize market. They can project themselves as innovative, customer centric brands offering specialized products and services thus improving customer trust and satisfaction.

The IoT ecosystem will have multiple stakeholders, diverse devices, platforms and service providers. Over the period it will evolve and bring in the digital strategies for enterprises. For insurers this means precise underwriting, identifying and accessing risks efficiently, processing claims effectively, harnessing powerful insights from data generated enabling better and faster decision making thus enhancing customer experience at every level. It is very important that insurers develop strategic plans to address technological transformations of IoT and its related advantages and risks while it's still evolving.